\documentclass[epj,draft]{svjour}

\def\be{\begin{equation}}
\def\ee{\end{equation}}

\begin{document}

\input epsf.sty

\title{Boundary critical behaviour of two-dimensional random Potts 
	models}
	\titlerunning{Random Potts 
	models}
	
\author{G\'abor Pal\'agyi\inst{1} \and Christophe Chatelain\inst{2}\and 
	Bertrand Berche\inst{2}\thanks{Authors for correspondence: 
	berche@lps.u-nancy.fr,} 
	\and Ferenc Igl\'oi\inst{3}$^,$\inst{4}\thanks{
	igloi@power.szfki.kfki.hu} }
	\authorrunning{G.Pal\'agyi, C. Chatelain,  B. Berche, and F. Igl\'oi}
	 
\institute{Department of Physics,
	University of Veszpr\'em, H-8201 Veszpr\'em, Hungary
	\and
	Laboratoire de Physique des
	mat\'eriaux (UMR CNRS No 7556), Universit\'e Henri Poincar\'e, Nancy 1,\\
 	F-54506 Vand\oe uvre les Nancy Cedex, France  
 	\and 
 	Research Institute for Solid State Physics and Optics, 
	H-1525 Budapest, P.O.Box 49, Hungary
	\and
	Institute for Theoretical Physics,
	Szeged University, H-6720 Szeged, Hungary
	}

\date{\today}


\abstract{
We consider random $q$-state Potts models for $3\le q \le 8$ on the square lattice
where the ferromagnetic couplings take two values $J_1>J_2$ with equal
probabilities. For any $q$ the model exhibits a continuous phase transition both in the
bulk and at the boundary. Using Monte Carlo techniques the surface and the 
bulk magnetizations are
studied close to the critical temperature and the critical exponents $\beta_1$ and
$\beta$ are determined. In the strip-like geometry the critical magnetization profile is
investigated with free-fixed spin boundary
condition and the characteristic
scaling dimension, $\beta_1/\nu$, is calculated from conformal
field theory. The critical exponents and scaling dimensions are found monotonously
increasing with $q$. Anomalous dimensions of the
relevant scaling fields are estimated and the multifractal behaviour at criticality
is also analyzed.
}

\maketitle

\newcommand{\centre}[2]{\multicolumn{#1}{c}{#2}} 
\newcommand{\crule}[1]{\multispan{#1}{\hrulefill}} 
\def\br{\noalign{\vskip2pt\hrule height1pt\vskip2pt}} 


\section{Introduction}\label{intro}

The presence of quenched i.e. time independent disorder could modify the cooperative
behaviour of physical systems with many degrees of freedom. In classical systems, where
thermal fluctuations dominate quantum fluctuations the effect of disorder in the
pure system phase-transition point can be analyzed by relevance-irrelevance criterions.
For second-order transitions, according to the well known Harris criterion~\cite{harris74},
disorder appears to be a relevant perturbation which moves the random system
towards a new fixed point  when the specific heat exponent
$\alpha$ of the pure system is positive. In the other situation, $\alpha<0$, the
disordered system remains in the pure model universality class. 
 The 
two-dimensional random-bond Ising
model (RBIM) corresponds to the marginal case.
It has been extensively studied in the 80's (for reviews 
of theoretical and numerical
studies, see Refs.~\cite{shalaev94} and \cite{selkeshchurtalapov94}, 
respectively).
The effect of randomness on first-order phase transitions
was considered later. Imry and Wortis 
argued that quenched disorder could induce a second-order phase 
transition~\cite{imrywortis79}. This argument was then rigorously proved
by Aizenman and Wehr, and Hui and 
Berker~\cite{aizenmanwehr89,huiberker89}: In two dimensions, even an 
infinitesimal amount of
quenched impurities changes the transition into a continuous one. 

The random bond
Potts model (RBPM) is the para\-digm of systems the pure version of
which undergoes a
second-order or a first-order transition, depending on the number of states, $q$,
per spin~\cite{wu82}. In two dimensions, the second-order regime $q\leq 4$ has been
considered by a number of authors, using perturbative field-theoretical
techniques~\cite{ludwig87,ludwigcardy87,ludwig90,dotsenkopiccopujol95a,jugshalaev,dotsenko95} or
Monte Carlo (MC) simulations~\cite{wisemandomany95,picco96,kim96}. On the other hand,
in the first-order regime, $q>4$, conformal perturbation techniques can not be used around the pure model
transition point and the resort to numerical calculations
becomes essential. Both Monte Carlo simulations and Transfer Matrix (TM) techniques, combined
to standard Finite Size 
Scaling (FSS)~\cite{chenferrenberglandau92,chenferrenberglandau95,chatelainberche98,yasar98,picco98,olsonyoung99} 
and conformal methods~\cite{picco97,cardyjacobsen97,jacobsencardy98,chatelainberche98b,chatelainberche99} 
were used at the 
random fixed point of self-dual disordered models to study bulk 
critical properties.

The surface properties of dilute or random-bond magnetic systems 
were on the other hand paid less attention. Generally surface quantities, such as
magnetization, energy-density, etc. are characterized by a different set of scaling
dimensions, than their bulk counterparts. For example in the pure Ising model, bulk
magnetization vanishes as $m \sim t^{\beta}$, with $\beta=1/8$, whereas for the surface
magnetization the  decay-law, $m_1 \sim t^{\beta_1}$, involves the surface exponent
$\beta_1=1/2$, where $t$ denotes the reduced temperature.
Quite generally, the scaling laws involving  surface and/or bulk exponents 
can be deduced
from the assumption of scale invariance. For example the singular part of the bulk,
$f_{\rm b}$, and surface, $f_{\rm s}$, free-energy densities in a $d$-dimensional system behave under
a scaling transformation, when lengths are rescaled by a factor $b>1$, $l'=l/b$, as
\begin{equation}
f_{\rm b}(t,h)=b^{-d}f_{\rm b}(b^{y_t}t,b^{y_h}h),
\label{eq-homogeneity}\end{equation}
\begin{equation}
f_{\rm s}(t,h,h_s)=b^{-(d-1)}f_{\rm s}(b^{y_t}t,b^{y_h}h,b^{y_{h_s}}h_s).
\label{eq-homogeneitys}\end{equation}
The whole set of bulk and surface critical exponents can be expressed in 
terms of the 
anomalous dimensions $y_i$ associated to the relevant 
scaling fields~\cite{binder83} (temperature $t$, bulk $h$ and boundary $h_s$
magnetic fields), for example $\beta=(d-y_h)/y_t$ and $\beta_1=(d-1-y_{h_s})/y_t$.

The $(1,1)$ surface of the disordered Ising model on a square lattice has recently been 
investigated through MC simulations by 
Selke {\it et al.}~\cite{selkeetal97,igloietal98}. (For a related study of
the critical behaviour at an internal defect line in the disordered Ising model,
see Ref.~\cite{szalmaigloi99}.)
The critical exponent $\beta_1$  was found robust against dilution 
keeping  the pure
system value $\beta_1=1/2$ and no logarithmic correction has been observed,
in contradistinction with the corresponding bulk behaviour. The surface properties of the 
8-state RBPM were also considered in 
Refs.~\cite{chatelainberche98,chatelainberche99}.

In this paper, we report  extensive MC and Transfer Matrix studies of the critical behaviour of both the
surface and bulk magnetizations of the disordered Potts ferromagnets for different
values of $3 \le q \le 8$. Our study extends previous investigations in several directions.
First, we investigate the temperature dependence of the bulk and surface magnetizations and
calculate the critical exponents $\beta$ and $\beta_1$. Second, we consider strip-like systems
with fixed spin boundary conditions (BC), determine the magnetization profile at the critical
temperature and calculate the scaling dimensions $x_b=2-y_h=\beta/\nu$ and $x_1=1-y_{h_s}=\beta_1/\nu$
from predictions of conformal invariance. Our third investigation concerns the possible
multifractal behaviour of the correlation function and the 
critical magnetization profile.
The $n$-th moments of both quantities are found to follow predictions of conformal invariance
and the scaling dimensions $x_b^{(n)}$ and $x_1^{(n)}$ are obtained $n$-dependent.

The structure of the paper is the following. In Section~\ref{model}, we 
present briefly the model and the simulation techniques. Section~\ref{off} is
devoted to the approach to criticality, while 
in Section~\ref{profile}, magnetization profiles in the transverse direction of
strips with fixed-free boundary conditions are computed. Multifractality is
studied in Section~\ref{Multi} and a discussion
of the results is given in Section~\ref{discussion}.

\section{Model and algorithms}\label{model}

\subsection{The random-bond Potts model}

We consider Potts-spin variables, $\sigma_{l,k}\in {1,2,\dots,q}$
on the sites of a square lattice with $l=1,2,\dots,L$ columns and
$k=1,2,\dots,K$ rows, with independent random nearest-neighbour ferromagnetic
interactions $J_{lk}$ and $J'_{lk}$ in the horizontal and vertical directions,
respectively,  which have the same distribution and could take two values,
$J_1>J_2$, with equal probabilities: 
\begin{equation}
	{\cal P}(J_{lk})=\frac{1}{2}\delta(J_{lk}-J_1)+
	\frac{1}{2}\delta(J_{lk}-J_2).
   	\label{probcoupl}
\end{equation}
The Hamiltonian of the model is thus written
\begin{equation}
   	-{\cal H}=\sum_{l,k}\left(J_{lk}\delta_{\sigma_{l,k},\sigma_{l+1,k}}+
   	J'_{lk}\delta_{\sigma_{l,k},\sigma_{l,k+1}}\right).
   	\label{Ham}
\end{equation}

In the thermodynamic limit $L,K \to \infty$ the model is self-dual
and the self-duality point
\be
[\exp(J_1/k_BT_c)-1][\exp(J_2/k_BT_c)-1]=q\;,
\label{duality}
\ee
corresponds to the critical point of the model if only one phase
transition takes place in the system. This assumption is strongly
supported by numerical calculations.

The degree of dilution in the system can be varied by changing the
ratio of the strong and weak couplings, $r=J_1/J_2$. At $r=1$, one
recovers the perfect $q$-state Potts model, whereas for $r \to \infty$
we are in the percolation limit, where $T_c=0$. The intermediate regime
of dilution $1<r<\infty$ is expected to be controlled by the random
fixed-point located at some $r=r^\star(q)$.

\subsection{Monte Carlo simulations}

For the simulation of spin systems, standard Metropolis algorithms based
on local updates of single spins suffer from the well known 
critical slowing down. As the second-order phase transition is approached, the
correlation length becomes longer and the system contains larger and larger clusters
in which all the spins are in the same state. Statistically independent
configurations can be obtained by local iteration rules only after a long
dynamical evolution which needs a huge number of MC steps and makes this
type of algorithm inefficient close to a critical point. 

Since disorder
changes the transition of the Potts model into a second-order one, the resort
to cluster update algorithms is more convenient~\cite{janke96,barkemanewmann97}. 
These algorithms are based on the
Fortuin-Kasteleyn representation~\cite{fortuinkasteleyn69} where
bond variables are introduced.
In the Swendsen-Wang algorithm \cite{swendsenwang87}, a 
cluster update sweep consists of three steps: 
Depending on the nearest neighbour 
exchange interactions, assign 
values to the bond variables, then identify clusters of spins connected by active
bonds, and eventually assign a random value to all the spins in a given cluster. 
The Wolff algorithm~\cite{wolff89} is a simpler variant in which only a single cluster
is flipped at a time. A spin is randomly chosen, then the cluster connected with
this spin is constructed and all the spins in the cluster are updated.

Both algorithms considerably improve the efficiency close to the critical
point and their performances are comparable in two dimensions, so in principle
one can equally choose either one of them. Nevertheless, when one uses particular boundary
conditions, with fixed spins along some surface for example, the Wolff algorithm
is less efficient, since close to criticality the unique cluster will often
reach the boundary and no update is made in this case.
In the following, we will consider two different series of simulations, one
with free BC where the Wolff algorithm will be used,
and the other with fixed-free BC for which we have chosen the
Swendsen-Wang algorithm.

\section{Approach to criticality}\label{off}

In this Section we consider square shaped systems, where $L$ and $K$
are equal, with $L$ ranging from $40$ to $640$ to check finite size
effects. In the vertical direction we impose periodic boundary
conditions, thus $\sigma_{l,K+1} \equiv \sigma_{l,1}$, for $l=1,2,\dots,L$,
whereas in the horizontal direction the boundary spins at $l=1$ and
$l=L$ are free. Thus we have a pair of $(0,1)$ surfaces, obtained
by cutting bonds along the vertical axis of the system.

 According to numerical studies
about the bulk quantities of the random model the finite size corrections
are very strong unless the calculations are 
performed close to the random fixed-point~\cite{chatelainberche99}.
The approximate position of $r^*(q)$ is listed in Table~\ref{table-beta} for different values of $q$,
as obtained from the maximum condition of the central charge of the
model~\cite{chatelainberche99,dotsenkojacobsenlewispicco98}. Our simulations
were performed at these fixed-point values of the dilution, but for comparison
we have also considered systems with somewhat different values of $r$.

We averaged over an ensemble of bond configurations and the number of
different realizations ranged from several hundreds to several thousands.
In the simulations the one-cluster flip Monte Carlo algorithm was used,
generating several $10^4$ clusters per realization close to the
critical point. As in earlier studies the statistical errors for each
realization were significantly smaller than those obtained by averaging
over the  different realizations. This is the reason of using a relatively
large number of realizations in the ensemble averaging.
The details of the parameters used in the MC simulations are given in
the case $q=8$, $r=10$ in Table~\ref{table-Wolff}.

\begin{table}
\small
\caption{Details of the parameters used for MC computations (Wolff
algorithm). These values are given for the case $q=8$, $r=10$. 
25\% of  cluster flips have been discarded for thermalization.}
\begin{center}
\vglue0mm
\begin{tabular}{@{}*{5}{l}}
\br
$t$ & $L\times K$  & \# of realizations & \# of cluster flips  \\
\hline
0.02 & 640$\times$320 &   317  & 10000 \\
0.05 & 160$\times$160 &   150    & 10000 \\
     & 320$\times$320 &   184    & 10000 \\
     & 640$\times$320 &   291    & 10000 \\
0.1  &   320$\times$320 &   303    & 10000 \\
0.15 &  320$\times$320 &   303    & 10000 \\
     & 160$\times$320 &   84    & 10000 \\
     & 160$\times$160 &   56    & 10000 \\
0.175 & 160$\times$160 &   100    & 10000 \\
0.2 &   320$\times$320 &   119    & 10000 \\
0.225 & 160$\times$160 &   100    & 5000 \\
0.25 &  160$\times$160 &   187    & 5000 \\
0.275 & 160$\times$160 &   100    & 5000 \\
0.3 &   160$\times$160 &   187    & 5000 \\
    & 80$\times$160 &   400    & 5000 \\
0.35 &  160$\times$160 &   187    & 5000 \\
0.4 &   160$\times$160 &   187    & 5000 \\
0.45 &  160$\times$160 &   187    & 5000 \\
0.5 &   160$\times$160 &  187    & 5000 \\
\br
\end{tabular}
\end{center}\label{table-Wolff}
\end{table}

In the MC simulations we calculated the magnetization profile, defined as
\be
[ m(l) ]_{\rm av}={1 \over K} [ | \sum_k m_{l,k} | ]_{\rm av}\;,
\label{ml}
\ee
where $m_{l,k}=(q \langle\delta_{\sigma_{l,k},1}\rangle -1)/(q-1)$ is the local Potts
order-parameter and the summation goes over the spins in the $l$-th
column, $k=1,2,\dots,K$. The brackets $\langle\dots\rangle$ and $[\dots]_{\rm av}$
stand for thermal and ensemble averages, respectively. 
The absolute values are taken in order to
obtain non-vanishing profiles for finite systems. The surface magnetization
is given by $[ m_1 ]_{\rm av}=[ m(1) ]_{\rm av}=[ m(L) ]_{\rm av}$.

\begin{figure}[ht]
\begin{center}
\vglue-0cm
\epsfxsize=8cm
\mbox{\epsfbox{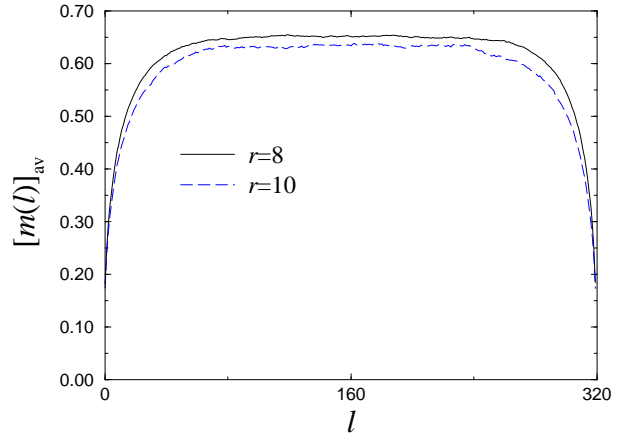}}
\end{center}\vskip-5mm
\caption{Profile of the $q=6$ model with free BC for two values of $r$ 
at a distance $t=0.02$ from the critical point.
The size is $320\times 320$ 
and average was performed over
180 and 44 realizations of disorder for $r=8$ and 10, respectively.}
\label{fig:Prof_q=6_r=10_K}
\end{figure}

The local magnetization, $[ m(l) ]_{\rm av}$, shows a monotonic decrease on approach
to the free surface, due to the reduced coordination number close to the
boundary. This is illustrated in Figure~\ref{fig:Prof_q=6_r=10_K} for  
the random $q=6$ model with
dilutions $r=8$ and $r=10$ at the same distance $t=0.02$ from the critical
temperature $T_c$ in equation~(\ref{duality}). The reduced 
temperature is
defined by $t=\mid K-K_c\mid / K$ where $K=J/k_BT$.
 As it can be seen, the randomness tends to reduce order, thus the
magnetization is decreasing with dilution. The magnetization profile
displays a plateau at the center of the system, the value of which defines
the bulk magnetization, $[ m_b]_{\rm av}=
[  m(L/2)]_{\rm av}$, at the given temperature. The surface
region of the profile has a characteristic size of $\xi_r$, which is
expected to scale like to the bulk correlation length, $\xi \sim t^{-\nu}$,
as the critical point is approached. For the random Potts model the
correlation length exponent, $\nu$, is close to $1$, for all values of
$q$~\cite{cardyjacobsen97}.

\begin{figure}[ht]
\begin{center}
\vglue-0cm
\epsfxsize=8cm
\mbox{\epsfbox{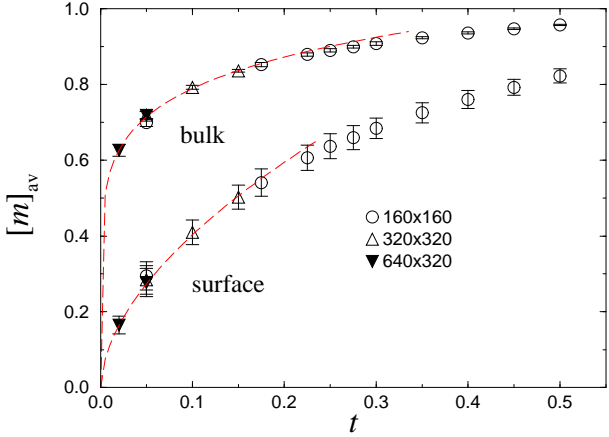}}
\end{center}\vskip-5mm
\caption{Temperature dependence of the surface and bulk magnetization
($q=8$, $r=10$). The dashed lines are guide for the eyes where the
extrapolated exponents of Table~\ref{table-beta} have been used.}
\label{fig:M-vs-t_q=8_r=10}
\end{figure}

In the thermodynamic limit, $L \to \infty$, as the critical temperature $T_c$
in (\ref{duality}) is approached, the magnetization profile $[  m(l) ]_{\rm av} $ goes to
zero as a power-law, $[  m(l) ]_{\rm av} \sim t^{\beta(l)}$, where $\beta(1)=\beta(L)=\beta_1$
and $\beta(l)=\beta$ for $\xi_r<l<L-\xi_r$, where $\beta_1$ and $\beta$ are the
usual surface and bulk critical exponents, respectively. 
The temperature dependence of bulk and surface magnetization is shown
in Figure~\ref{fig:M-vs-t_q=8_r=10}.
To estimate the values
of these critical exponents from simulation data one may define 
temperature-dependent effective exponents
\be
\beta_{\rm eff}(l)=\frac{{\rm d} \ln[  m(l) ]_{\rm av}}
	{{\rm d} \ln t}\;,
\ee
which are approximated by using data at discrete temperatures, say, $t+\Delta t/2$
and $t-\Delta t/2$. In the limit of sufficiently small $\Delta t$ and $t$ the
effective exponents approach the true critical exponents, presuming that the
system is large enough so that finite-size effects play no role (to avoid
finite-size effects, $L$ should be much larger than the size of the surface
region, $\xi_r$, and the bulk correlation length, $\xi$).

\begin{figure}[ht]
\begin{center}
\vglue-0cm
\epsfxsize=8cm
\mbox{\epsfbox{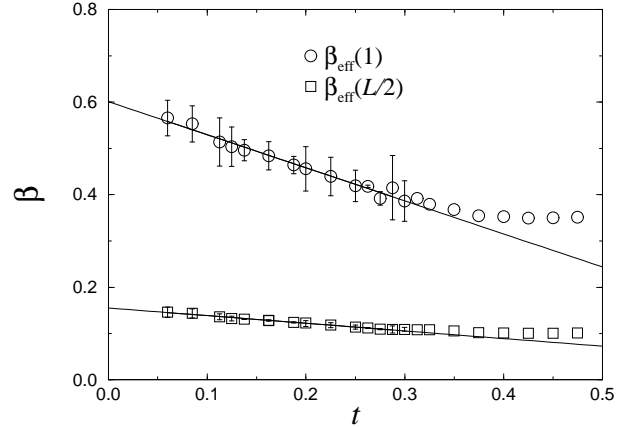}}
\end{center}\vskip-5mm
\caption{Temperature-dependent effective exponents
$\beta_{\rm eff}(l)$ for the surface and bulk magnetization
for $q=8$, $r=10$. In the case of the bulk, the error bars are smaller than
the symbol sizes.}
\label{fig:Effective-expst_q=8_r=10}
\end{figure}

\begin{figure}[ht]
\begin{center}
\vglue-0cm
\epsfxsize=8cm
\mbox{\epsfbox{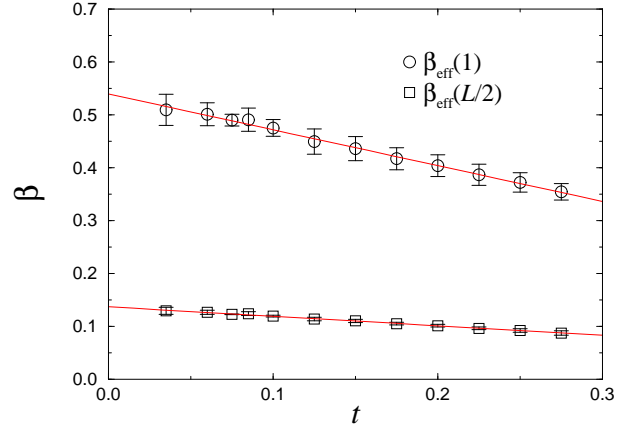}}
\end{center}\vskip-5mm
\caption{Same as 
	Figure~
	\ref{fig:Effective-expst_q=8_r=10}
	,
	for $q=3$, $r=5$.}
\label{fig:Effective-expst_q=3_r=5}
\end{figure}

In the actual calculation, we approached the critical point by calculating 
$\beta_{\rm eff}(l)$
for several temperatures, $t$, ranging from $0.05$ to $0.45$ with $\Delta t=0.05$.
As shown in Figure~\ref{fig:Effective-expst_q=8_r=10} and 
\ref{fig:Effective-expst_q=3_r=5},
the effective  exponents of the random $q=8$ and $q=3$ Potts models approach
linearly their limiting value. To obtain an accurate estimate for the true critical
exponents we analyzed and extrapolated the data for $\beta_{\rm eff}(l)$ using different
types of correction to scaling forms.
The most successful correction for the surface magnetization, written as 
\be
	[  m_1 ]_{\rm av}\sim at^{\beta_1}(1+bt^\theta),
	\label{eq-cooection}
\ee
 was obtained  
with $\theta\simeq 1$.

The estimates for the surface and bulk magnetization
critical exponents are given for different dilutions in
Table~\ref{table-beta-r}. The pure case values at $q=3$ and 4 
have also been computed
to check the method~\cite{carlonigloi98}. The case $q=4$ involves known 
logarithmic
corrections which were taken into account~\cite{carlonigloi98,salassokal97}. 
In the random case, as already observed in Finite Size Scaling
studies by different authors
(e.g. in Ref.~\cite{picco98}), due to crossover effects, the disorder 
amplitude has a sensible influence on the exponents. At the optimal disorder amplitude
deduced from the behaviour of the central charge~\cite{chatelainberche99}, the
exponents reach their random fixed point values 
 summarized in Table~\ref{table-beta}. 
 As seen in  the Table, both  surface and bulk 
critical exponents depend on the value of
$q$ and there is a monotonic increase with increasing $q$. 
Provided that the correlation length exponent is close to 1 for any
value of $q$, this observation is in
accordance with previous estimates 
on the bulk magnetization scaling dimension
$x_b=\beta/\nu$ obtained in Ref.~\cite{chatelainberche99} at the
 random fixed point and recalled in the Table.

\begin{table}
\small
\caption{Bulk and surface exponents deduced from the approach to criticality
at different disorder amplitudes. The variation of the exponents, outside the
standard deviation, is due to crossover effects.}
\begin{center}
\vglue0mm
\begin{tabular}{@{}*{7}{l}}
\br
$q$ & $r$ & $\beta_1$ & $\Delta\beta_1$ & $\beta$ & $\Delta\beta$  \\
\hline
3 & 1 & 0.541 & 0.009 & 0.112 & 0.002  \\
  & 4 & 0.542 & 0.010 & 0.135 & 0.010 \\
  & 5 & 0.542  & 0.011 & 0.1361 & 0.0008 \\
  & 10& 0.504 & 0.020 & 0.141 & 0.004\\
\hline
4 & 1 & 0.666 & 0.009 & 0.0831 & 0.0002  \\
  & 4 & 0.56 & 0.02 & 0.1332  & 0.0004 \\
  & 7& 0.561 & 0.022 & 0.142   & 0.002\\
  & 10& 0.534 & 0.029 & 0.146   & 0.003\\
\hline
6 & 8 & 0.581 & 0.028 & 0.149   & 0.003 \\
  & 10 & 0.566 & 0.018 & 0.149  & 0.003 \\
\hline
8 & 10 & 0.597 & 0.023 & 0.1513  & 0.0004 \\
\br
\end{tabular}
\end{center}\label{table-beta-r}
\end{table}

\begin{table}
\small
\caption{Bulk and surface exponents deduced from the approach to criticality
at the $q$-dependent optimal disorder amplitude $r^\star$. The last column
recalls the bulk scaling dimension $x_b=\beta/\nu$ obtained with the same
disorder amplitudes in Ref.~
\cite{chatelainberche99}
.}
\begin{center}
\vglue0mm
\begin{tabular}{@{}*{6}{l}}
\br
$q$ & $r^\star$ & $\beta_1$ & $\beta$ & $x_b$  \\
\hline
3 & 5 & 0.542(10) & 0.136(1) & 0.132(3)  \\
4 & 7 & 0.561(22) & 0.142(2) & 0.139(3) \\
6 & 8 & 0.581(28) & 0.149(3)  & 0.146(3)\\
8 & 10 & 0.597(23) & 0.151(1) & 0.150(3) \\
\br
\end{tabular}
\end{center}\label{table-beta}
\end{table}

At this point we are going to check the self-averaging properties of the local
magnetization in the vicinity of the system critical temperature. For non-self-averaging
quantities, the reduced variance does not vanish in the thermodynamic limit, indicating that 
fluctuations never become negligible~\cite{aharonyharris96}. Here we studied different
moments of the local magnetization and determined the corresponding critical exponent, $\beta^{(n)}$,
defined through $[  m^n ]_{\rm av}^{1/n}
\sim t^{\beta^{(n)}}$~\cite{derrida84,wisemandomany98a,wisemandomany98}. 
For self-averaging quantities, $\beta^{(n)}$ is expected to be independent
of $n$. As seen in Table~\ref{table-self} the critical exponents both for the bulk
and surface magnetizations are found to be independent of $n$, at least within the error of the
numerical calculations. Thus we conclude that the local magnetization as the critical point is
approached is self-averaging. This observation is in agreement with the expectation, that outside
the critical point, where the system size is much larger than
the correlation length, the central limit theorem is expected to apply, which
implies self-averaging behaviour. At the critical point, however, where the above
argument does not hold one may obtain non-self-averaging behaviour, as was observed
recently by Olson and Young~\cite{olsonyoung99} for the bulk
spin-spin correlation function. We are going to study this issue in Section 5.

\begin{table}
\small
\caption{Test of self-averaging in the off-critical behaviour of surface and bulk 
magnetization
($q=8$, $r=10$): The power law behaviours of 
$[  m_1^n ]_{\rm av}^{1/n}$ and 
$[  m_b^n ]_{\rm av}^{1/n}$ define only two exponents, for boundary 
and bulk behaviours, respectively.}
\begin{center}
\vglue0mm
\begin{tabular}{@{}*{4}{l}}
\br
$n$ &  $\beta_1^{(n)}$ & $\beta^{(n)}$  \\
\hline
0.01 &  0.601(21) & 0.1516(4)  \\
1    &  0.597(23) & 0.1513(4) \\
2    &  0.592(21) & 0.1513(5)  \\
3    &  0.587(20) & 0.1513(5)  \\
4    &  0.582(21) & 0.1513(5)  \\
\br
\end{tabular}
\end{center}\label{table-self}
\end{table}

\section{Magnetization profile in strips at criticality}\label{profile}

\subsection{Conformal profiles in homogeneous strips}

In a system which is geometrically constrained by the presence of surfaces, 
the local order-parameter
evolves from the surface towards the bulk behaviour and the
appropriate way to describe the position-dependent physical quantities
is to use density profiles rather than bulk and surface
observables. This is particularly important close to the critical point
where the correlation length, which measures the surface region, is diverging.

\begin{figure}[ht]
\begin{center}
\vglue-0cm
\epsfxsize=8cm
\mbox{\epsfbox{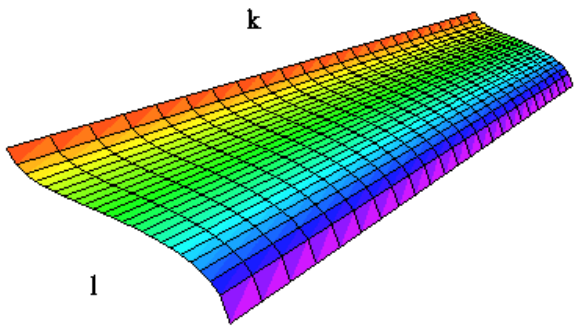}}
\end{center}\vskip-5mm
\caption{Profile with fixed-free BC in the transverse direction.
In the computations, periodic BC are used in the longitudinal direction. 
}
\label{fig:3d_Fxd-fr}
\end{figure}

For example, in a homogeneous critical system, infinite in one direction,
$k\in ]-\infty,+\infty[$,
and confined between two parallel plates, 
which are at a
large, but finite distance $L$ apart, the local order parameter 
$m(l)$
 varies
with the distance $l\in [1,L]$ from one of the plates as a smooth function of
$l/L$. According to the Fisher and de
Gennes scaling theory~\cite{fisherdegennes78}:
\be
m(l)=L^{-x_{b}}F_{ab} (l/L)\;,
\label{fisher_degennes}
\ee
where  $a$ and $b$ denotes the boundary conditions at the two plates.
In the middle of the strip, $l=L/2$, one recovers the Finite Size Scaling
behaviour of the bulk magnetization $m(L/2)\sim L ^{-x_{b}}$.
In two-dimensions, conformal invariance gives further constraints on the
profile: Considering a semi-infinite system described by
$z=x+{\rm i}y=\rho {\rm e}^{{\rm i}\theta}$, $y\ge 0$,  
with boundary conditions $a$ and $b$ on the positive
and negative $x$ axis, respectively, under the logarithmic transformation
$w(z)=\frac{L}{\pi}\ln z=k+{\rm i}l$, one obtaines the above
strip geometry. Ordinary scaling, as in Eq.~(\ref{fisher_degennes}) then 
implies a functional form
 in the half-plane~\cite{burkhardtxue91}:
 \be
 m(z)=y^{-x_b}G_{ab}(x/\rho),
 \label{m-halfab}
 \ee
which is transformed in the strip geometry to the following expression
\begin{eqnarray}
&m(w)=|w'(z)|^{-x_b}m(z)\nonumber\\ 
&=\left[{L\over \pi} \sin\left(  \pi{l \over L}\right) 
\right]^{-x_{b}}
G_{ab} (\cos \pi l/L)\;,\nonumber\\ 
\label{confprof}
\end{eqnarray}
%
%
where the scaling function $G_{ab} (\cos \pi l/L)$ depends on the universality class of
the model and on the type of the boundary conditions at the two edges of the strip.
In the following, we
consider the fixed-free BC, i.e. we fix the
spins to the state $\sigma_{1,k}=1$ only at one boundary of the system, the other
surface being free. As in Section~\ref{off}, we choose 
periodic BC
in the vertical direction, $\sigma_{l,K+1}=\sigma_{l,1}$. The conformal and scaling
results are strictly valid as $K \to \infty$, however the corrections for $K\gg L$ are
expected to be small. We indicate 
these  boundary conditions by setting $a=1$, $b=f$ in equation~(\ref{confprof}).
For such conformally
invariant, non-symmetric boundary conditions, the scaling function has been
predicted for several models~\cite{burkhardtxue91,carlonigloi98}:
\be
G_{1f}(\cos \pi l/L)  = {\cal A} \left[ \cos \left( \frac{\pi l}{2 L} \right) 
\right]^{x_1}.
\label{mag+f}
\ee
We mention that with the functional form in equation (\ref{mag+f}) one recovers
the usual finite-size scaling behaviour, $m(L)\sim L^{-x_1}$, close to the free surface
at criticality. The typical shape of the magnetization profile in a strip with
fixed-free BC is shown in Fig.~\ref{fig:3d_Fxd-fr}.

\subsection{Application to random systems}

In random systems, the conformal invariance prescriptions are
recovered after disorder average~\cite{chatelainberche98b,igloirieger98}. 
The bulk critical properties of disordered Potts models have been investigated
at criticality through conformal techniques by various 
authors~\cite{picco97,cardyjacobsen97,jacobsencardy98,chatelainberche98b,chatelainberche99},
using the longitudinal correlation function decay along periodic strips or the
magnetization profiles in square-shaped systems with fixed boundary
conditions along all the surfaces. Here, we consider the
magnetization profiles of long strips with fixed-free
boundary conditions in the transverse direction. Since a sufficient strip width
is needed in order to apply the continuum limit conformal results 
in the transverse direction, 
TM techniques are useless (the strip width is limited to $L\le 10$ using
the connectivity TM) and MC simulations are preferred (with $L\le 40$).
The parameters used in this work
for the MC simulations are given in Table~\ref{table-parameters}.
We mention that in spite of the large sizes used, the continuum
limit is only approximately reached and
perturbing effects will be expected close to the boundaries.

\begin{table}
\small
\caption{Parameters for the MC simulations at criticality. The same parameters are used
for all values of $q$. The number of disorder realizations is increased at
larger sizes. At each disorder realization, 1000 MC sweeps are discarded 
and 5000 MC sweeps are used
to compute physical quantities.}
\begin{center}
\vglue0mm
\begin{tabular}{@{}*{5}{l}}
\br
$L$ & $K$ & \# of realizations  \\
\hline
10 & 100, 200, 300, 400 and 500 & 1000 \\
14 & 100, 200, 300, 400 and 500 & 1000 \\
18 & 138, 277, 555 and 833 & 1000 \\
24 & 80 and 96 & 1000 \\
   & 192, 384 and 500 & 4000 \\
30 & 100 and 200 & 1000 \\
   & 300, 400 and 500 & 4000 \\
40 & 100 and 200 & 1000 \\
   & 300, 400 and 500 & 4000 \\
\br
\end{tabular}
\end{center}\label{table-parameters}
\end{table}

\begin{figure}[ht]
\begin{center}
\vglue-0cm
\epsfxsize=8cm
\mbox{\epsfbox{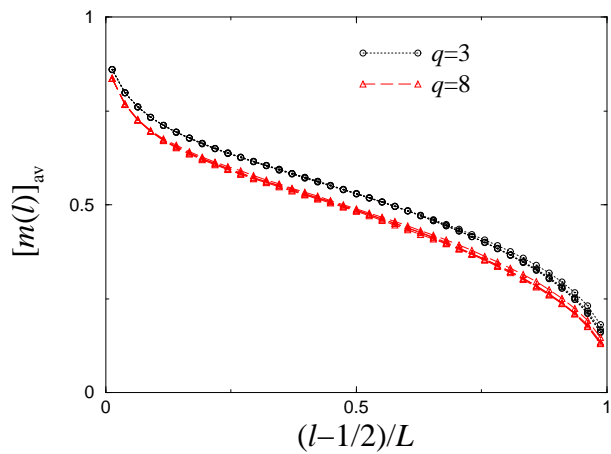}}
\end{center}\vskip-5mm
\caption{Profile with fixed-free BC for two values of $q$.
The size $L\times K$ of the strips of width $L=40$ ranges between $40\times 200$ 
(upper profiles) and $40\times 500$ (lower profiles), 
and average was performed over
1000 to 4000 realizations of disorder. Error bars are smaller than the symbols. 
}
\label{fig:Prof_Fxd-fr}
\end{figure}

Examples of profiles with fixed-free BC are shown in 
Figure~\ref{fig:Prof_Fxd-fr} for $q=3$ and 8 for different strip lengths $K$ at a
fixed width $L=40$. The influence of the length $K$ of the strip
becomes negligible when $K\geq 300$.

Introducing the variable $\zeta =\frac{l-1/2}{L}$ in equations~(\ref{confprof}) and 
(\ref{mag+f}),
one thus expects the following behaviour:
\be
	[  m(\zeta ) ]_{\rm av}={\cal A}(\pi/L)^{x_b}\times
	[\sin\pi \zeta ]^{-x_b}\times [\cos\pi \zeta /2]^{x_1}.
	\label{profz}
\ee
In order to simplify the following expressions, we introduce the 
functions 
$f(\zeta )=\sin\pi \zeta $ and $g(\zeta )=\cos\pi \zeta /2$, and
the ratio 
\be
	R(\zeta ,\zeta ')=\frac{[  m(\zeta ) ]_{\rm av}}
	{[  m(\zeta ') ]_{\rm av}}.
	\label{ratio}
\ee
Since $f(\zeta )$ is symmetric with respect to the middle of the strip 
$\zeta =\frac{1}{2}$, the surface dimension $x_1$ can be deduced from local symmetric
values of the profile:
\be
	R(\zeta ,1-\zeta )=\left[\frac{g(\zeta )}{g(1-\zeta )}\right]^{x_1}
	=[\cot\pi \zeta /2]^{x_1},
	\label{Rx1}
\ee
or 
\be	
	x_1=\frac{\ln R(\zeta ,1-\zeta )}{\ln [\cot\pi \zeta /2]}.
	\label{x1}
\ee

\begin{figure}[ht]
\begin{center}
\vglue-0cm
\epsfxsize=8cm
\mbox{\epsfbox{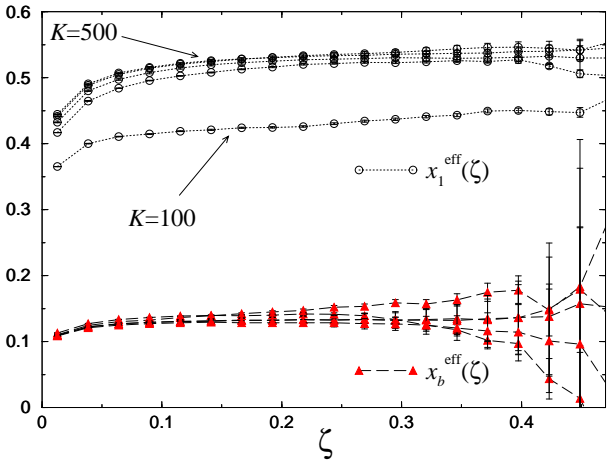}}
\end{center}\vskip-5mm
\caption{Examples of effective surface and bulk scaling dimensions for $q=4$
and a disorder amplitude $r^\star=7$.
The size of the strip are $40\times K$, with $K=100$, 200, \dots, 500.
Between $K=300$ and 500, the effective exponents remain constant up to the
accuracy of the computations. 
Average was performed over 1000 to
4000 realizations of disorder.}
\label{fig:Exple_expsts_q=4_40}
\end{figure}

Examples of effective surface scaling dimensions, according to 
Equation~(\ref{x1}), are shown
in Figure~\ref{fig:Exple_expsts_q=4_40} for $q=4$ with 4000 different 
configurations of couplings for strips of width $L=40$ and increasing
lengths from $K=100$ to 500.

\begin{figure}[ht]
\begin{center}
\vglue-0cm
\epsfxsize=8cm
\mbox{\epsfbox{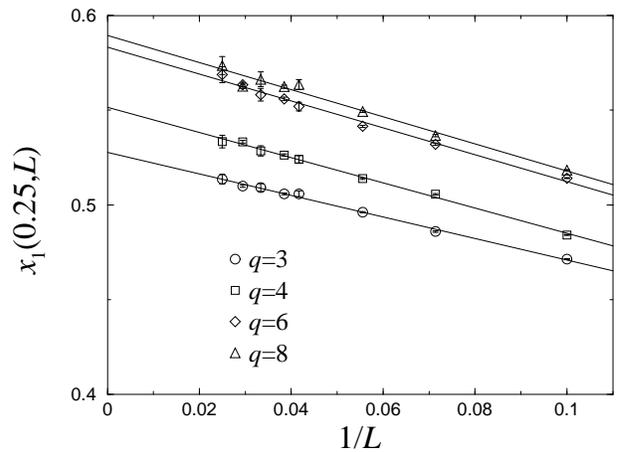}}
\end{center}\vskip-5mm
\caption{Effective surface exponent at $\zeta =0.25$ plotted against $L^{-1}$.
The intercept corresponds to the 
extrapolated value in the
thermodynamic limit: $x_1(0.25,\infty)$.}
\label{Expsts_x1-vs-L_q=4}
\end{figure}

In the thermodynamic limit, these scaling dimensions should be unambiguously
determined for any value of the position $\zeta $ in the transverse direction of the
strip. In practice, due to the finite size of the system and to lattice effects
close to the boundaries~\footnote[1]{Close to the surfaces, $\zeta \to 0$, lattice 
effects and probable corrections to scaling spoil the results, while
near the middle of the strip width, $\zeta \to 1/2$, the precision becomes very 
low due to the 
proximity of the points used for the computation
of the exponents.}, equation~(\ref{x1})
defines effective quantities $x_1(\zeta ,L)$ 
which do depend on the position $\zeta $ along the strip, and also 
on the strip width and
evolve towards the right limit when $L\to\infty$. 
As it was already visible in Figure~\ref{fig:Prof_Fxd-fr}, 
between $K=300$ and $K=500$, the strip length 
can be considered to
be long enough in order to avoid finite-size effects in the long direction.
For these strip lengths, one can also observe a plateau in 
Figure~\ref{fig:Exple_expsts_q=4_40} where there is no significant 
variation of the
the effective exponents which
remain almost constant in the region $\zeta =0.20-0.35$, and whose extrapolated 
values
should be consistent.
The  values computed in the plateau region are 
thus studied as $L$ increases and extrapolation is
made towards the thermodynamic limit $L\to\infty$.
The dependence of the effective exponents on the strip width is shown in
Figure~\ref{Expsts_x1-vs-L_q=4} for $\zeta =0.25$. A linear extrapolation 
leads to the  estimations of $x_1(\zeta ,\infty)$, 
collected in Table~\ref{table-x_z}.
 The results exhibit a good stability 
relative to the position $\zeta $ and allow a final determination of the
scaling dimension $x_1$, given in the last column of the table.
On the other hand, we  observe important corrections to 
scaling in the boundary behaviour, since the surface exponent falls down
rapidly as $\zeta \to 0$. This effect is the possible origin of the discrepancy with
the boundary scaling dimension found in Ref.~\cite{chatelainberche98}
for $q=8$: $x_1\simeq 0.47$, which is recovered here in the limit $\zeta \to 0$.

\begin{table}
\small
\caption{Extrapolation in the thermodynamic limit of the 
 scaling dimension of the surface magnetization, $x_1(\zeta ,\infty)$, 
 measured
at different values of $\zeta $. The last column presents our definitive
determination for each value of $q$.}
\begin{center}
\vglue0mm
\begin{tabular}{@{}*{8}{l}}
\br
&&\centre{4}{$\zeta $}\\
&&\crule{4}\\
$q$ & $r^\star$ & $\zeta \to 0$ & 0.20 & 0.25 & 0.30  & $x_1$ \\
\hline
3 & 5   & 0.438(1)   & 0.518(1)  & 0.526(2) & 0.525(3) & 0.523(2)  \\
4 & 7   & 0.453(1)   & 0.541(2)  & 0.553(2) & 0.552(3) & 0.549(2) \\
6 & 8   & 0.478(1)   & 0.567(2)  & 0.577(2) & 0.574(4) & 0.573(3) \\
8 & 10  & 0.482(1)   & 0.577(2)  & 0.588(3) & 0.588(5) & 0.584(3) \\
\br
\end{tabular}
\end{center}\label{table-x_z}
\end{table}

\begin{figure}[ht] 
\begin{center}
\vglue-0cm
\epsfxsize=8cm
\mbox{\epsfbox{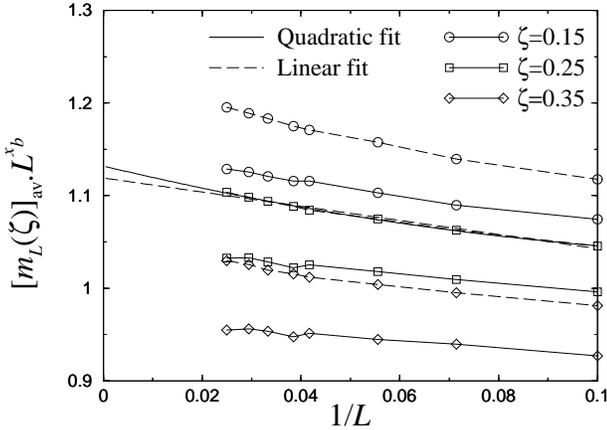}}
\end{center}\vskip-5mm
\caption{Extrapolation to infinite width of the profiles. The solid lines corresponds
to $q=8$ and the dashed lines to $q=3$. Two fitting curves are presented
(for $\zeta=0.25$, $q=3$), a
least square linear fit (dotted line) and quadratic (solid line). The quadratic 
fit is more
accurate.}
\label{Extrap_prof}
\end{figure}

For the bulk exponent, using the quantity
\be
	R(\zeta ,1/2)=[f(\zeta )]^{-x_b}[\sqrt 2 g(\zeta )]^{x_1},
\label{eqR}
\ee
we can form a combination where $x_1$ cancels, leading to:
\begin{eqnarray}
	x_b=&\{\ln R(\zeta ,1/2)\times \ln [\sqrt 2 g(1-\zeta )]\nonumber\\ 
	&-\ln R(1-\zeta ,1/2)\times \ln [\sqrt 2 g(\zeta )]\}\nonumber\\ 
	&\times\left\{\ln [f(\zeta )]\times \ln\left[\frac{g(\zeta )}{g(1-\zeta )}\right]
	\right\}^{-1}. \nonumber\\ 
	\label{xb}
\end{eqnarray}
This expression involves three points of the profile and is thus subject to
stronger numerical fluctuations than the surface scaling dimension,
 as shown in 
Figure~\ref{fig:Exple_expsts_q=4_40}. Unfortunately, compared to previous precise
determinations, no accurate
extrapolation can be made here. One can however 
check the conformal expression~(\ref{profz}). For that
purpose, it is necessary to extrapolate the rescaled profiles $[  m_L(\zeta )]
\times L^{x_b}$,
 obtained at finite
sizes, towards the thermodynamic limit. Considering, as we did before, that the 
longer strips are large enough to be unaffected by finite-size effects in the 
long direction, extrapolation to infinite width $L$ only will be done.
For each position $\zeta $ in the transverse direction, we plot the corresponding
local magnetization as computed for different strip widths and 
then extrapolate in the limit $L\to\infty$.
 Examples of linear and
quadratic least square fits are shown in Figure~\ref{Extrap_prof}.
The latter one has been preferred. 

\begin{figure}[ht] 
\begin{center}
\vglue-0cm
\epsfxsize=8cm
\mbox{\epsfbox{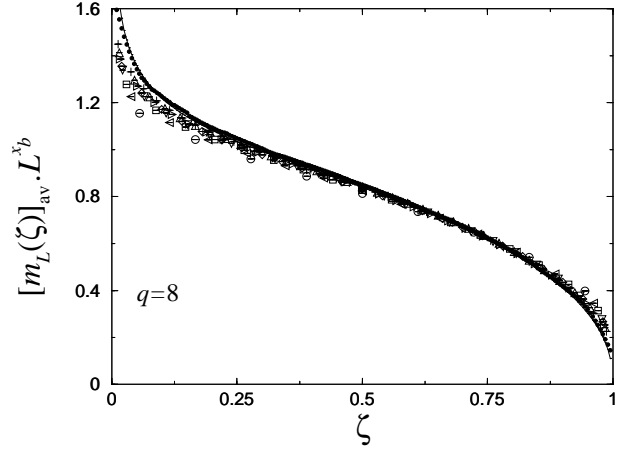}}
\end{center}\vskip-5mm
\caption{Rescaled profiles at different strip widths for $q=8$ (open symbols) 
and extrapolation
to $L\to\infty$ (full circles). The solid line is the conformal expression 
where the
 amplitude is the only free parameter. The agreement between the extrapolated data
 and the conformal profile is very good.}
\label{Profil_Extrap_q=8}
\end{figure}

It leads to an  extrapolated profile, which
is compared to 
 the conformal expression
 in Figure~\ref{Profil_Extrap_q=8} for $q=8$. 
 In the conformal formula,
the scaling dimensions of Tables~\ref{table-beta} and \ref{table-x_z} are 
entered and the amplitude
 is the only free parameter. In spite of the small strip widths considered, 
 the extrapolation procedure
 introduced in Fig.~\ref{Extrap_prof} is very accurate, since the agreement
 between extrapolated data (full circles) and Eq.~\ref{profz} (solid line)
  is fairly satisfactory. This is a strong evidence which 
  supports the validity of the conformal expression
 for the profile.

\section{Multifractal behaviour at criticality}\label{Multi}

In this section, we study the possible multifractal behaviour at the critical point
and consider different moments for the magnetization profile, $[m^n(l)]^{1/n}_{\rm av}$,
and that of the correlation function $[G_{\sigma}^n(r)]^{1/n}_{\rm av}$.
The characteristic exponents, $x_b^{(n)}$ and $x_1^{(n)}$, in the bulk and at the
surface, respectively, are expected to vary with $n$ for multifractal behaviour. Indeed,
it was first Ludwig~\cite{ludwig87} who predicted multifractality in the bulk
correlation function of the random bond Potts model by 
conformal perturbative methods
(see also the results by Lewis in Refs.~\cite{lewis98,lewis99}), which was confirmed
recently by Olson and Young~\cite{olsonyoung99} by MC simulations in the square geometry.
Here we rather work in the strip geometry and calculate both the bulk and surface scaling
exponents.

The calculations about the moments of the critical profile are parallel with that in the
previous section and the scaling dimensions are extracted from the expected functional
form:
\be
	[  m^n(\zeta ) ]_{\rm av}^{1/n}={\cal A}(\pi/L)^{x_b^{(n)}}\times
	[\sin\pi \zeta ]^{-x_b^{(n)}}\times [\cos\pi \zeta /2]^{x_1^{(n)}}\;,
	\label{profzmoment}
\ee
which is analogous to (\ref{profz}) for the average behaviour, i.e. for $n=1$. We
note that the {\it typical} behaviour corresponds to $n=0$,i.e.
$\exp [ \ln m(\zeta ) ]_{\rm av}$.

\begin{figure}[ht] 
\begin{center}
\vglue-0cm
\epsfxsize=8cm
\mbox{\epsfbox{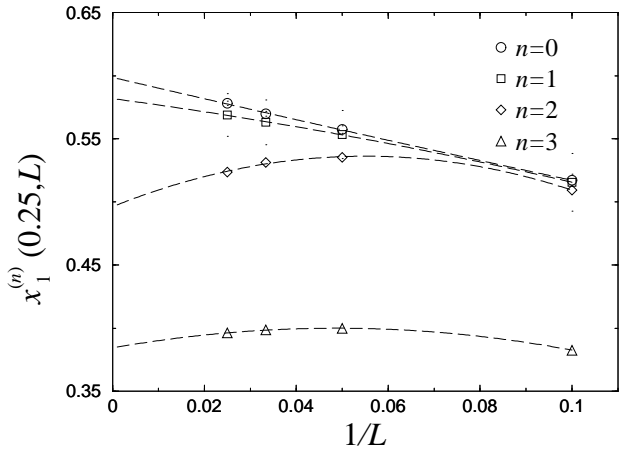}}
\end{center}\vskip-5mm
\caption{Effective surface dimensions of the $n^{\rm th}$-order moments of the 
magnetization profile, evaluated at $\zeta =0.25$ for $q=8$ and $n=0$, 1, 2, 3.
Dashed lines denote results of quadratic extrapolation.}
\label{MF_surf_q=8}
\end{figure}

In the actual calculation we have considered the $q=8$ model on strip-like samples
with $K=500$ and $L=10,20,30$ and $40$ and the average is performed over $5000$
realizations. Using the method of the previous section first we deduced from equation
(\ref{profzmoment}) the effective, size and position dependent surface scaling
dimensions, $x_1^{(n)}(\zeta,L)$, which are then extrapolated to the thermodynamic
limit, $L \to \infty$. The extrapolation procedure is demonstrated in 
Figure~\ref{MF_surf_q=8},
whereas the extrapolated data are presented in Table~\ref{C}. As one can see in 
this Table
the critical point surface magnetization of the random bond Potts model shows multifractal
behaviour: The scaling dimensions of the different moments of the surface magnetization,
$x_1^{(n)}$, are monotonously decreasing with $n$. Note that for $n=2$ the surface scaling
dimension is very close to the pure (and random) Ising value of $x_1=1/2$.

\begin{table}
\small
\caption{Extrapolation in the thermodynamic limit of the 
 scaling dimension associated to the moments of the surface 
 magnetization and measured
at different values of $\zeta $ ($q=8$). The last row presents our definitive
determination.}
\begin{center}
\vglue0mm
\begin{tabular}{@{}*{8}{l}}
\br
&\centre{4}{$\zeta$}\\
&\crule{4}\\
$n $ & $0.20$ & $0.25$ & $0.30$ & $0.35$ & $x_1^{(n)}$  \\
\hline
0 &  0.601(7)  &  0.600(7)   &  0.602(9)  & 0.599(11) & 0.600(9)   \\
1 &  0.585(5)  &  0.582(6)   &  0.582(8)  & 0.579(11) & 0.582(8)   \\
2 &  0.496(3)  &  0.496(4)   &  0.500(6)  & 0.498(8) &  0.498(5)  \\
3 &  0.387(11) &  0.384(14)  &  0.384(18)  & 0.381(25)&  0.384(17)  \\
\br
\end{tabular}
\end{center}\label{C}
\end{table}


Next we turn to study the multifractal behaviour of the bulk magnetization. As
mentioned in the previous section, from the magnetization profiles in finite strips one
cannot extract precise estimates for the scaling dimension $x_b^{(n)}$. Therefore we
used a different technique, based on the Bl\"ote and Nightingale connectivity 
transfer matrix~\cite{blotenightingale82}. Since 
transfer operators in the time direction do not commute in disordered systems,
the free energy density is defined by the  leading Lyapunov
exponent. For an infinitely long strip of width $L$
with periodic boundary conditions, the 
leading Lyapunov exponent is given by the Furstenberg
method~\cite{furstenberg63}: 
\begin{equation}
	\Lambda_0(L)=\lim_{m\to\infty}\frac{1}{m}
	\ln\left|\!\left|\left(\prod_{j=1}^m 
	{\bf T}_j\right)
	\mid\! v_0\rangle\right|\!\right|,
\label{eq-Furst}
\end{equation}	
where ${\bf T}_j$ is the transfer matrix and
$\mid\! v_0\rangle$ is a unit initial vector. The  
free energy density
is thus given by $[f_0(L)]_{\rm av}=-L^{-1}\Lambda_0(L)$.
For a specific disorder realization, the spin-spin correlation function
along the strip
\begin{equation}
 G_{\sigma}(k)=\frac{q\langle\delta_{
\sigma_{l,1}\sigma_{l,k+1}}\rangle-1}{q-1},
\label{eq-Gu}
\end{equation}
follows from the application of products of transfer matrices on the
ground state eigenvector associated to $\Lambda_0$ (for details see, e.g. 
Ref.~\cite{chatelainberche99}).

\begin{figure}[ht]
\begin{center}
\vglue-0cm
\epsfxsize=6.5cm
\mbox{\epsfbox{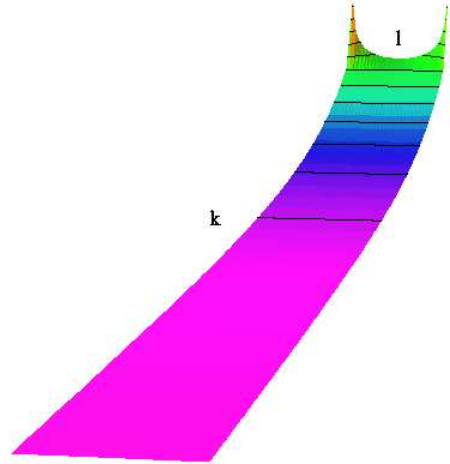}}
\end{center}\vskip0mm
\caption{Spin-spin correlation function on a long strip with
periodic BC in the transverse direction (the upper left corner also 
corresponds to the upper right corner). This Figure is a 3D plot 
where the value of the correlation function
between the origin (upper left corner) 
and any lattice site $(k,l)$ is plotted on a vertical axis 
at position $(k,l)$.
}
\label{fig:3d_PBC}
\end{figure}

We will now assume that
conformal covariance can be applied to the order parameter
correlation function and its moments. 
In the infinite
complex plane $z=x+{\rm i}y$ 
  the correlation function exhibits the usual
algebraic decay at the critical point
$
	[ G_\sigma^n(\rho)]_{\rm av}^{1/n}={\rm const}
	\times 
	\rho^{-2x_b^{(n)}}, 
$
where
$\rho=\mid\! z_1-z_2\!\mid$. Under the logarithmic transformation
$w=\frac{L}{2\pi}\ln z=k+{\rm i}l$ , one gets the usual  exponential decay along the 
strip
$
[{G}_\sigma^n(k)]_{\rm av}^{1/n}={\rm const}\times
\exp\left(-\frac{2\pi}{L}x_b^{(n)} k\right)$ 
(see Fig.~\ref{fig:3d_PBC} for an illustration).
 The scaling 
dimension $x_b^{(n)}$ can thus be deduced from an exponential fit. 

This method was used in Ref.~\cite{chatelainberche99} for the average correlation
function, i.e. for $n=1$. In this work, we calculate the higher moments, as well as
the typical behaviour, which corresponds to $n=0$.

\begin{figure}[ht] 
\begin{center}
\vglue-0cm
\epsfxsize=8cm
\mbox{\epsfbox{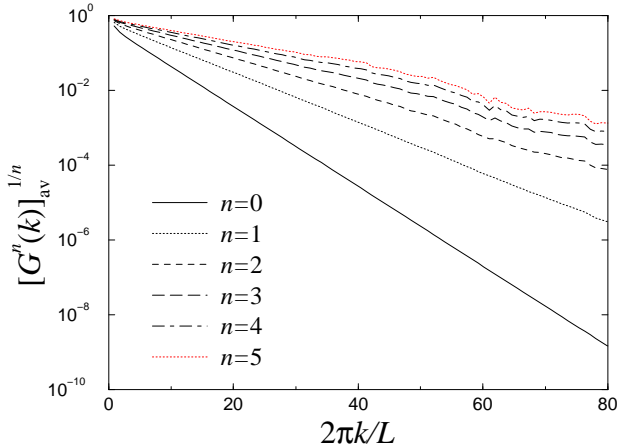}}
\end{center}\vskip-5mm
\caption{Moments of the spin-spin correlation function (semi-log scale) for $q=8$, $L=7$.}
\label{Moments_G_q=8}
\end{figure}

For each strip size ($L=2-8$),  we considered systems of length $\sim 10^6$
and an average is performed over $80\times 10^3$ disorder configurations. 
For a given strip the effective, size-dependent exponents follow from a linear fit in a semi-log 
plot, as exemplified in Figure~\ref{Moments_G_q=8}.
It is clearly seen that the scaling dimensions are different for the
different moments of the spin-spin correlation function. The effective exponents are then
extrapolated as $L\to\infty$ and the
estimated
values are presented in 
Table~\ref{table-multixb} for different values of $q$ (at $r^\star(q)$).

\begin{table}[ht] 
\small
\caption{Scaling dimensions of the moments of the 
spin-spin correlation function computed at $r=r^\star$.}
\begin{center}
\vglue0mm
\begin{tabular}{@{}*{6}{l}}
\br
&\centre{4}{$x_b^{(n)}$}\\
&\crule{4}\\
$n$ & $q=3$ & $q=4$ & $q=6$ & $q=8$   \\
\hline
0  & 0.154(1) & 0.177(1) & 0.207(1) & 0.234(1)   \\
1  & 0.132(1) & 0.138(1) & 0.146(1) & 0.150(1)   \\  
2  & 0.116(1) & 0.114(1) & 0.114(1) & 0.112(1)   \\  
3  & 0.104(1) & 0.097(1) & 0.094(2) & 0.091(2)   \\  
4  & 0.095(2) & 0.087(2) & 0.081(2) & 0.077(2)   \\
5  & 0.088(2) & 0.079(2) & 0.072(2) & 0.068(2)   \\
\br
\end{tabular}
\end{center}\label{table-multixb}
\end{table}

Again the critical point bulk magnetization shows multiscaling behaviour, for all
$q \ge 3$ the scaling dimensions, $x_b^{(n)}$, are monotonously 
decreasing with $n$. One
can make a comparison with perturbative results:
\be
x_b^{(n)}=x_b^p-\frac{9}{32}(n-1)\left[
\frac{2}{3}\epsilon +(A+B(n-2))\epsilon^2+O(\epsilon^3)\right],
\label{ludwig}
\ee
where $x_b^p$ is the bulk exponent in the pure model, $\epsilon$ is
a small expansion parameter related to 
the deviation of the pure model's
central charge to that of the Ising model ($\epsilon=2/15$ and 
$1/3$ for $q=3$ and 4,
respectively~\cite{lewis99}), and $A=11/12-4\ln 2$ and $B=\frac{1}{24}(33-29\sqrt{3}\pi/3)$ 
are numerical 
factors. The first-order  term in  Eq.~(\ref{ludwig}) is due to
Ludwig~\cite{ludwig87} and the second-order term to Lewis~\cite{lewis98}.
This result is known to be valid close to $q=2$ and for the lower moments. The
comparison with our data is made in Table~\ref{tableludwig}. 
One can observe a good agreement
with the result of Lewis  for $q=3$, especially for the lower moments.
For $q=4$, the perturbation expansion probably breaks down already at $n=3$,
since $x_b^{(3)}$ is found to be larger than
$x_b^{(2)}$ with the formula of Lewis. It is remarkable that the second-order
term in the perturbation expansion vanishes at $n=2$, which could
explain the stability of the
exponent associated to the second moment, with respect to variations of $q$, has already
been noticed by Olson and Young~\cite{olsonyoung99}. 


\begin{table}[ht] 
\small
\caption{Scaling dimensions of the moments of the 
spin-spin correlation function computed at  $r=r^\star$. The numerical estimates
on the first column are compared to the results of the 1st and 2nd order perturbation
theory in Eq.~(\ref{ludwig}).}
\begin{center}
\vglue0mm
\begin{tabular}{@{}*{5}{l}}
\br
&\centre{4}{$x_b^{(n)}$ for the 3-state Potts model}\\
&\crule{4}\\
$n$ & num. & 1st order & 2nd order   \\
\hline
0  & 0.154(1) & 0.158 & 0.157    \\
2  & 0.116(1) & 0.108 & 0.118    \\  
3  & 0.104(1) & 0.083 & 0.110   \\
\hline  
&\centre{4}{$x_b^{(n)}$ for the 4-state Potts model}\\
&\crule{4}\\
$n$ & num. & 1st order & 2nd order   \\
\hline
0  & 0.177(1) & 0.188 & 0.181   \\
2  & 0.114(1) & 0.063 & 0.120   \\  
3  & 0.097(1) & 0.000 & 0.167  \\
\br
\end{tabular}
\end{center}\label{tableludwig}
\end{table}

\begin{figure}[ht] 
\begin{center}
\vglue-0cm
\epsfxsize=8cm
\mbox{\epsfbox{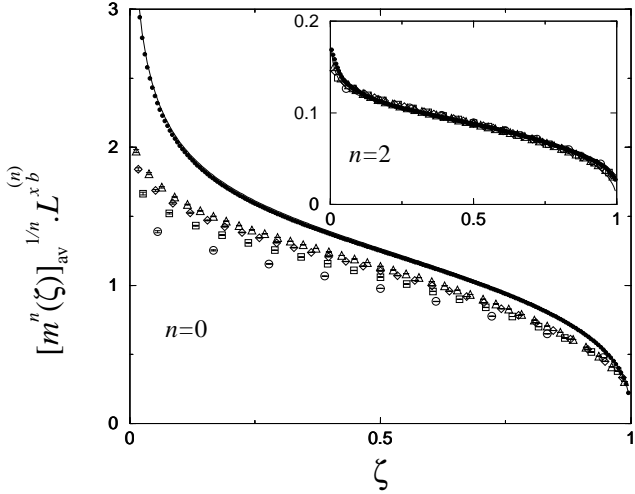}}
\end{center}\vskip-5mm
\caption{Rescaled $n^{\rm th}$-order profiles at different 
strip widths for $q=8$ (open symbols) 
and extrapolation
to $L\to\infty$ (full dots). The solid line is the 
conformal expression with only
the amplitude as a free parameter.}
\label{Profil_MF-n=0-et-2}
\end{figure}

We close this section by presenting different moments of the magnetization profile,
$[m^n(l)]_{\rm av}^{1/n}$, as extrapolated towards $L \to \infty$. In Figure~\ref{Profil_MF-n=0-et-2} two
moments ($n=0$ and $n=2$) of the scaled profiles are plotted for the $q=8$ model, where
the open symbols represents the finite-width data. The extrapolated profile, denoted by
full circles is in perfect agreement with the conformal results in equation (\ref{profzmoment}),
where the scaling dimensions $x_b^{(n)}$ and $x_1^{(n)}$ are taken from 
Tables~\ref{table-multixb} and \ref{table-y},
respectively, and the only fitting parameter is the amplitude in equation (\ref{profzmoment}).
We consider this agreement as a strong evidence in favour of the validity of the
conformal expression for the averaged moments of the order parameter profile.

\section{Discussion}\label{discussion}

In this paper the critical behaviour of the two-dimensional random bond Potts model is
studied by MC simulations, transfer matrix techniques and conformal methods. New
features of our present work are the following. {\it i)} For the first time we have
investigated the surface critical behaviour of the model and determined the surface
magnetization critical exponent, $\beta_1$, from approach to criticality. In addition
we got estimates for the corresponding bulk exponent, $\beta$. {\it ii)} We have studied
the critical point magnetization profiles in strip-like geometries and deduced the scaling
dimension $x_1=\beta_1/\nu$ from the predictions of conformal invariance. {\it iii)} We
have presented numerical evidence for the multifractal behaviour of the surface and
bulk magnetizations at the critical point and the scaling dimensions of the averaged
moments, $x_b^{(n)}$ and $x_1^{(n)}$ are calculated. {\it iv)} Finally, we have demonstrated
that the different moments of the critical magnetization profiles, as well as the
correlation function obey conformal invariance.

The critical exponents and the scaling dimensions of the average quantities are
continuously varying functions of $q$, however an exponent relation $4 \beta=\beta_1$
is approximately satisfied, as can be observed in Table~\ref{table-beta}. The anomalous dimensions of the
relevant scaling fields in equations (\ref{eq-homogeneity}) and (\ref{eq-homogeneitys})
can be estimated using the scaling relations:
$y_t=1/\nu=x_b/\beta=x_1/\beta_1$, $y_h=2-x_b$, and $y_{h_s}=1-x_1$. Their
values are presented in Table~\ref{table-y}. While $y_t$ remains close to 1
(but $\nu$ satisfies the limit $\nu\geq 2/d$~\cite{chayesetal}) 
for all values of $q$, the anomalous dimensions related to the magnetic field
vary with $q$.

\begin{table}[ht] 
\small
\caption{Anomalous dimensions of the relevant bulk and surface scaling fields.}
\begin{center}
\vglue0mm
\begin{tabular}{@{}*{7}{l}}
\br
$q$ & $r^\star$ & $x_b/\beta$ & $x_1/\beta_1$ & $2-x_b$ & $1-x_1$  \\
\hline
3 & 5 & 0.971(29) & 0.965(6) & 1.868(3) & 0.477(2)   \\
4 & 7 & 0.979(35) & 0.979(13) & 1.862(3) & 0.451(2)   \\
6 & 8 & 0.980(40) & 0.986(18) & 1.854(3) & 0.427(3)   \\
8 & 10 & 0.993(26)& 0.978(18) & 1.849(3) & 0.416(3)   \\
\br
\end{tabular}
\end{center}\label{table-y}
\end{table}


\begin{acknowledgement}

We thank M.A. Lewis for stimulating discussions.
This work has been supported by the French-Hungarian
cooperation program "Balaton" (Minis\-t\`ere des Af\-faires 
Etran\-g\`eres-\-O.M.F.B.), the Hungarian
National Research Fund under grants No OTKA TO23642, TO25139 and
OTKA M 028418 and by the Ministry of Education under grant No. FKFP
0596/1999. The Laboratoire de Physique des Mat\'eriaux is Unit\'e Mixte
de Recherche CNRS No 7556.
This
work was supported by computational facilities: Centre 
Charles Hermite in Nancy, and  CNUSC in Montpellier under project No C990011. 
\end{acknowledgement}


\end{document}